\documentclass[11pt]{article}
\usepackage{venice}
\usepackage{epsfig}
\usepackage{bm}

\begin{document}

\title{Nuclear structure and  double beta decay}

\author{Petr Vogel}
\address{ 
Kellogg Radiation Laboratory, Caltech,
Pasadena, California, 91125, USA.}

\abstract
{Study of the neutrinoless double beta decay, $0\nu\beta\beta$, includes a variety
of problems of nuclear structure theory. They are reviewed here. The problems range
from the mechanism of the decay, i.e. exchange of the light Majorana neutrino neutrino
vesus the exchange of some heavy, so far unobserved particle. Next, the proper
expressions for the corresponding operator are described
that should include the effects of the nucleon
size and of the recoil order terms in the hadronic current. The issue of proper treament
of the short range correlations, in particular for the case of the heavy particle
exchange, is discussed also. The variety of methods employed these days in the
theoretical evaluation of the nuclear matrix elements $M^{0\nu}$ is briefly
described and the difficulties causing the spread and hence uncertainty in the values
of $M^{0\nu}$ are discussed. Finally, the issue of the axial current quenching, and of 
the resonance enhancement in the case of double electron capture are described.  
}

\section{Introduction}
Study of the neutrinoless double beta decay is the most sensitive current test
of the total lepton number conservation. If observed, it would serve as a proof that 
not only the lepton number $L_{tot}$ is not  conserved, but also that
neutrinos are massive Majorana fermions. Hence the study of this process is one
of the important parts of the search for ``Physics Beyond the Standard Model''. 
Different aspects of the complex of problems associated with the $\beta\beta$ decay are
described in this focus issue. Since the process involves nuclei, i.e. complicated
many body systems, the analysis of the decay necessary involves various nuclear
structure problems. In this article the general questions of nuclear structure,
relevant for the understanding of the $\beta\beta$ decay rate, are discussed.
In other works in this issue the particular approximate techniques for the
treatment of the corresponding nuclear matrix elements are described. Here I try
to discuss the more general framework. Various aspects, both theoretical and experimental,
of the $\beta\beta$ decay have been reviewed many times. I quote here
just some of the review articles\cite{SuhCiv98,EV02,AEE07,Gom11,Rod11}, earlier 
references can be found there.  

In double beta decay two neutrons, bound in the ground state of an even-even initial 
(or parent) nucleus
are transformed into two bound protons, typically also in the the ground state of the
final (or granddaughter) even-even nucleus, with the simultaneous emission of two
electrons only for the $0\nu\beta\beta$ mode, or two electrons plus two $\bar{\nu}_e$
for the $2\nu\beta\beta$ mode. Transitions leading to the excited bound states
of the final nucleus are sometimes kinematically allowed as well; however, 
we will concentrate here
on the most often studied case of the ground state to ground state decays
\begin{equation}
(Z,A)_{g.s.} \rightarrow (Z+2,A)_{g.s.} + 2e^- + (2\bar{\nu}_e) ~.
\label{eq:1}
\end{equation}

Note that transitions of two bound protons into two bound neutrons (nuclear charge is decreased
by two units in that case) with the emission of either two positrons, or a positron accompanied by the 
electron capture, or by two simultaneous electron captures, are also kinematically
allowed for several even-even nuclei, again in both two-neutrino and neutrinoless modes.
The lepton phase space for these transitions is suppressed compared to the decays
of the type (\ref{eq:1}), while the nuclear structure issues are analogous. Hence, again
we will concentrate on the decays described in Eq. (\ref{eq:1}). However, the $0\nu ECEC$
mode in few cases can have a resonance character ($Q \rightarrow 0$). 
The possible enhancement in those special cases is described in the Appendix 

The $\beta\beta$ decay, in either mode, can proceed only if the initial nucleus is stable against
the standard $\beta$ decay (both $\beta^-$ and $\beta^+$ or $EC$). That happens exclusively
in even-even nuclei, where moreover the ground state is always $I^{\pi} = 0^+$.
The $\beta\beta$ decay rate is a steep function of the energy carried by the 
outgoing leptons (i.e. of the decay $Q$-value). Hence, transitions with larger $Q$-value
are easier to observe. For this reason 
in Table \ref{tab:1} I list all candidate nuclei with Q values larger than 2 MeV that are
particularly well suited for the study of the $\beta\beta$ decay.

There has been a significant progress recently in the accuracy of the atomic mass determination using 
various trap arrangements. In many cases the $Q$-values are determined with accuracy
better than 1 keV, making the search for the all important $0\nu\beta\beta$ decay
mode easier; in Table \ref{tab:1} these more recent $Q$-value determinations are shown,
together with the corresponding references. 

In both modes of the  $\beta\beta$ decay the rate can be expressed as a product
of independent factors that depend on the atomic physics (the so called phase-space 
factors $G^{0\nu}$ and $G^{2\nu}$) that include
also the $Q$-value dependence as well as the fundamental physics constants, nuclear structure (the nuclear
matrix elements $M^{0\nu}$ and $M^{2\nu}$), and for the $0\nu\beta\beta$ mode the possible
particle physics parameters (the effective neutrino mass $\langle m_{\beta\beta} \rangle$ in the
simplest case). Thus
\begin{equation}
\frac{1}{T_{1/2}^{0\nu}} = G^{0\nu} |M^{0\nu}|^2 |\langle m_{\beta\beta} \rangle|^2 ~;
\hspace{1cm} \frac{1}{T_{1/2}^{2\nu}} =  G^{2\nu} |M^{2\nu}|^2 ~.
\label{eq:rate}
\end{equation} 

\begin{table}[htb]
\begin{center}
    \caption{Candidate nuclei for $\beta\beta$ decay with $Q > 2$ MeV } 
\label{tab:1}
\medskip
\begin{tabular}{|l|lccc|}\hline
Transition & $Q$-value & Ref. &    $(G^{2\nu})^{-1}$ &  $(G^{0\nu})^{-1}$  \\ 
  & (keV) & & (y $\times$ MeV$^{-2}$) &  (y $\times$ eV$^2$) \\ \hline
$^{48}_{20}$Ca $\rightarrow$ $^{48}_{22}$Ti & 4273.6$\pm$ 4  & \cite{web} & 9.7 $\times 10^{16}$ & 4.1$\times 10^{24}$ \\
$^{76}_{32}$Ge  $\rightarrow$ $^{76}_{34}$Se & 2039.006 $\pm$ 0.050 & \cite{Q-76} & 2.9$\times 10^{19}$ & 4.1$\times 10^{25}$ \\
$^{82}_{34}$Se  $\rightarrow$ $^{76}_{36}$Kr & 2995.50 $\pm$ 1.87 &  \cite{web} & 8.8$\times 10^{17}$ & 9.3$\times 10^{24}$ \\
$^{96}_{40}$Zr  $\rightarrow$ $^{96}_{42}$Mo & 3347.7 $\pm$ 2.2 & \cite{web} & 2.0$\times 10^{17}$ & 4.5$\times 10^{24}$ \\
$^{100}_{42}$Mo  $\rightarrow$ $^{96}_{44}$Ru & 3034.40 $\pm$ 0.17 & \cite{Q-100} & 4.1$\times 10^{17}$ & 5.7$\times 10^{24}$ \\
$^{110}_{46}$Pd  $\rightarrow$ $^{96}_{48}$Cd & 2017.85 $\pm$ 0.64 & \cite{Q-110}  & 9.6$\times 10^{18}$ & 5.7$\times 10^{25}$ \\
$^{116}_{48}$Cd  $\rightarrow$ $^{116}_{50}$Sn & 2813.50 $\pm$ 0.13 & \cite{Q-116} & 4.8$\times 10^{17}$ & 5.3$\times 10^{24}$ \\
$^{124}_{50}$Sn $\rightarrow$ $^{124}_{52}$Te & 2287.80 $\pm$ 1.52 & \cite{web} & 2.3$\times 10^{18}$ & 9.5$\times 10^{24}$ \\
$^{130}_{52}$Te  $\rightarrow$ $^{130}_{54}$Xe & 2527.01 $\pm$ 0.32 & \cite{Q-130} & 8.0$\times 10^{17}$ & 5.9$\times 10^{24}$ \\ 
$^{136}_{54}$Xe $\rightarrow$ $^{136}_{56}$Ba  & 2458.7 $\pm$ 0.6 & \cite{Q-136} & 7.9$\times 10^{17}$ & 5.5$\times 10^{24}$ \\ 
$^{150}_{60}$Nd  $\rightarrow$ $^{150}_{62}$Sm & 3371.38 $\pm$ 0.20 & \cite{Q-150} & 3.2$\times 10^{16}$ & 1.3$\times 10^{24}$ \\
\hline
\end{tabular}
\end{center}
\end{table}

The values of $G^{0\nu}$ and $G^{2\nu}$ are also listed in Table \ref{tab:1}. The entries there are taken from
Ref. \cite{BV}, and are not corrected for the small changes in $Q$ and $g_A$ since that time. Also, since by
convention the nuclear matrix elements  $M^{0\nu}$ are dimensionless,
the nuclear radius appears in them as a multiplicative factor. To compensate for it, 
the phase-space factor  $G^{0\nu}$ is
proportional to $R^{-2}$, where $R=r_0 \times A^{1/3}$ is the nuclear radius. In  Table \ref{tab:1} the value
$r_0$ = 1.2 fm was used. (Note that, obviously, the values of the phase-space factors depend on the convention
used for $r_0$ and $g_A$. One has to keep that issue in mind when using the Eq. (\ref{eq:rate}) to relate
the half-lifes and nuclear matrix elements (see e.g. \cite{Cowell, Grab10}).)

\begin{figure}[htb]
\begin{center}
\includegraphics[scale=0.40]{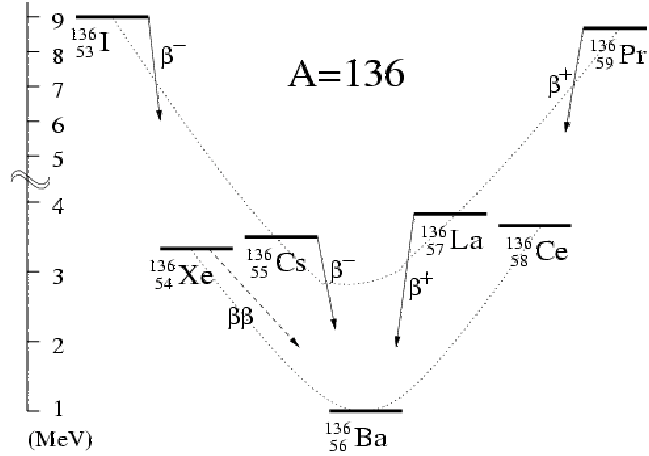}
\end{center}
\caption{ Atomic masses of the isotopes with $A$ = 136. Nuclei $^{136}$Xe, $^{136}$Ba
and $^{136}$Ce are stable against the ordinary $\beta$ decay; hence they exist in nature.
However, energy conservation alone allows the transition
$^{136}$Xe $\rightarrow$ $^{136}$Ba + $2e^-$ (+ possibly
other neutral light particles) and the analogous decay of $^{136}$Ce with the positron
 emission. }
\label{fig_bb}
\end{figure}

Double beta  transitions are
possible and potentially observable because
nuclei with even $Z$ and $N$ are more bound than the odd-odd nuclei with
the same $A = N + Z$.  A typical example is shown in Fig. \ref{fig_bb}. With one
exception, all nuclei in Table \ref{tab:1} have an analogous mass pattern. The one
exception is $^{48}$Ca where the intermediate nucleus $^{48}$Sc can be in principle
reached by the $\beta^-$ decay of $^{48}$Ca with $Q$= 278 keV. However, the ground 
state of $^{48}$Sc is $6^+$ and the first excited state at 131 keV is $5^+$. $\beta$
decays with a large nuclear spin change are heavily suppressed; in this particular
case the $\beta^-$ decay of $^{48}$Ca has not been observed as yet, while the 
$2\nu\beta\beta$ decay has been observed.

The two-neutrino mode ($2\nu\beta\beta$)
is just an ordinary beta decay of two bound neutrons occurring simultaneously since
the sequential decays are forbidden by the energy conservation law.
For this mode, clearly, the lepton number is conserved and
this decay is allowed in the standard model of electroweak interaction.
It has been repeatedly observed in a number of cases and proceeds with a typical half-life
of $\sim 10^{19-20}$years for the nuclei with $Q$-values above 2 MeV.
In contrast, the neutrinoless
mode ($0\nu\beta\beta$) obviously violates the law
of lepton number conservation and is forbidden in the standard model. Hence, its observation
would, as already stated, be a signal of the `` physics beyond the standard model".

The two modes of the $\beta\beta$ decay have some common and some distinct features.
The common features are:
\begin{itemize}
\item  The leptons carry essentially all available energy. The nuclear recoil is negligible,
$Q/Am_p \ll 1$.
\item The transition involves the $0^+$ ground state of the initial nucleus and
(in almost all cases) the $0^+$ ground state of the final nucleus. In few cases the
transition to an excited $0^+$ or $2^+$ state in the final nucleus is energetically possible, but
suppressed by the smaller phase space available. (But the $2\nu\beta\beta$ decay to
the excited $0^+$ state has been observed in few cases.)
\item Both processes are of second order of weak interactions, $\sim G_F^4$, hence inherently
slow.  The phase space consideration alone (for the $2\nu\beta\beta$ mode $\sim Q^{11}$ and
for the $0\nu\beta\beta$ mode $\sim Q^5$) give preference to the $0\nu\beta\beta$ which is,
however, forbidden by the lepton number conservation and therefore much slower (at least by a 
factor $10^5$ for the nuclei listed in Table \ref{tab:1}) than the $2\nu\beta\beta$ decay.
\end{itemize}
The distinct features are:
\begin{itemize}
\item In the $2\nu\beta\beta$ mode the two neutrons undergoing the transition are uncorrelated
(but decay simultaneously) while in the $0\nu\beta\beta$ the two neutrons are correlated.
In the  $2\nu\beta\beta$ mode the corresponding momentum transfer $q$ is restricted by
the decay $Q$-value; hence $qR \ll 1$. On the other hand, in the $0\nu\beta\beta$ mode the
momentum transfer is of the order of the nucleon Fermi momentum $q \sim q_{Fermi}$;
hence $qR \ge 1$ in that case.
\item In the $2\nu\beta\beta$ mode the sum electron kinetic energy $T_1 + T_2$
spectrum is continuous and peaked below $Q/2$.
This is due to the electron masses and the Coulomb attraction.
As $T_1 + T_2 \rightarrow Q$ the
spectrum approaches zero approximately like $(\Delta E/Q)^6$.
\item On the other hand
 in the $0\nu\beta\beta$ mode the sum of the electron kinetic
 energies is fixed, $T_1 + T_2 = Q$, smeared only by the detector resolution. This  
allows one to separate the two modes experimentally
by measuring the sum energy of the emitted electrons with a good energy
resolution, even if the decay rate for the $0\nu\beta\beta$ mode is much smaller
than for the   $2\nu\beta\beta$ mode.
\end{itemize}
 
Another hypothetical mode of double-beta decay is often considered in the literature,
the decay accompanied by Majoron emission. Majorons are supposed to be very light or massless
particles $\chi^0$ that couple to neutrinos. A variety of approaches involving massless
(or almost massless) scalar particles and their impact
on $0\nu\beta\beta$ decay  have been considered. The discussion of this topic goes beyond
the scope of the present review; a rather complete list of references can be found e.g. in Ref. 
\cite{Rod11}. In all of these hypotheses the sum electron spectra are continuous with the
characteriostic shape (neglecticng Coulomb effects)
\begin{equation}
\frac{d\Gamma^{0\nu M}}{dE_1 dE_2} \sim (Q - E_1 -_2)^n p_1 p_2 E_1 E_2 ~,
\end{equation}
where the ``index'' $n = 1 - 7$ depends on the model in question. The decay
rate is expressed as
\begin{equation}
\Gamma^{0\nu M} = | \langle g_{\chi} \rangle |^2 |M_{\chi}|^2 G_{\chi} (Q, Z) ~,
\end{equation}
where $ \langle g_{\chi} \rangle$ is the model dependent averaged coupling constant. 
For $n = 1$ that constant is experimentally constrained to be less than about $10^{-5}$.

 \section{Mechanism of the $0\nu\beta\beta$ decay}
 
The relation between the $0\nu\beta\beta$-decay
rate and the effective Majorana mass $\langle m_{\beta\beta} \rangle$
is to some extent problematic. The rather conservative
assumption leading to Eq.(\ref{eq:rate}) is that the only possible way
the $0\nu\beta\beta$ decay can occur is through the exchange
of a virtual light, but massive,
Majorana neutrino between the two nucleons undergoing the transition,
and that these neutrinos interact by the standard left-handed weak currents. But that is not
the only theoretically possible mechanism. 
Lepton number violating (LNV) interactions involving so far unobserved
much heavier ($\sim$ TeV) particles
might lead to a comparable $0\nu\beta\beta$ decay rate.

In general $0\nu\beta\beta$ decay can be generated by (i) light massive Majorana
neutrino exchange or (ii) heavy particle exchange (see, e.g. Refs.\cite{heavy,Pre03}),
resulting from LNV dynamics at some scale $\Lambda$ above the electroweak one.
The relative size of heavy ($A_H$) versus light
particle ($A_L$) exchange contributions to the decay amplitude
can be crudely estimated as follows~\cite{Mohapatra:1998ye}:
\begin{equation}
A_L \sim G_F^2  \frac{\langle m_{\beta \beta} \rangle}{\langle k^2 \rangle}  ,~
 A_H \sim G_F^2  \frac{M_W^4}{\Lambda^5}  ,~
\frac{A_H}{A_L} \sim \frac{M_W^4 \langle k^2 \rangle }
{\Lambda^5  \langle m_{\beta \beta} \rangle }  \ ,
\label{eq_estimate}
\end{equation}
where $\langle m_{\beta \beta} \rangle$ is the effective neutrino
Majorana mass,
$\langle k^2 \rangle \sim ( 100 \ {\rm MeV} )^2 $ is the
typical light neutrino virtuality, and $\Lambda$ is the heavy
scale relevant to the LNV dynamics.
Therefore,  $A_H/A_L \sim O(1)$ for  $\langle m_{\beta \beta} \rangle \sim 0.1-0.5$
eV and $\Lambda \sim 1$ TeV, and  thus the LNV dynamics at the TeV
scale leads to similar $0 \nu \beta \beta$-decay rate as the
exchange of light Majorana neutrinos with the effective mass
$\langle m_{\beta \beta} \rangle \sim 0.1-0.5$ eV.

Obviously, the $0\nu\beta\beta$ lifetime measurement by itself
does not provide the means for determining the underlying mechanism.
The spin-flip and non-flip exchange can be, at least in principle,
distinguished by the measurement of the single-electron spectra or
polarization.  However, in most cases the
mechanism of light Majorana neutrino exchange, and of
heavy particle exchange cannot be separated by the observation
of the emitted electrons. We will not discuss here the
possible ways of determining which mechanism is responsible
for the transition, once the  $0\nu\beta\beta$ decay has been observed.
However, obviously, the corresponding nuclear matrix elements $M^{0\nu}$
depend on that.

 Lets comment now on the
main differences between the mechanism involving light or heavy
particle exchange.
We will show later that the evaluation of the nuclear matrix element
 $M^{0\nu}$ can be performed in the closure approximation, i.e.
without explicit treatment of the virtual states
in the intermediate odd-odd nucleus. Thus 
\begin{equation}
M^{0\nu} = \langle f || O^K || i \rangle~,      
\end{equation}   
where the  operator $O^K$  creates two protons and  annihilate two neutrons.
In addition, the operator  $O^K$ depends on the distance between these nucleons,
and on their other quantum numbers. The main difference between the mechanism 
involving the light massive Majorana neutrino exchange and the heavy ($\sim$ TeV)
particle exchange is the range of the operator  $O^K$. The light neutrino
exchange represents two point-like vertices separated by the distance $r \sim 1/q$.
The decay rate is then proportional to the square of the effective Majorana
neutrino mass as in Eq. (\ref{eq:rate}). On the other hand
the heavy particle exchange represents a single point-like vertex
(six fermions, four hadrons and two leptons), i.e. dimension 9 operator. 
Proper treatment of the short range nucleon-nucleon repulsion is obviously crucial
in that case. The relation between the
neutrino mass and the decay rate is not simple in that case.

Independently of its mechanism the existence of $0\nu\beta\beta$ decay
would mean that on the elementary particle level a six fermion
lepton number violating amplitude
transforming two $d$ quarks into two $u$ quarks plus two electrons
is nonvanishing.  As was first pointed out by Schechter and Valle\cite{SV82},
already thirty years ago,
this fact alone would guarantee that neutrinos are massive
Majorana fermions. This qualitative statement (or theorem),
however, as we pointed out above, does not in general allow one 
to deduce the magnitude of the neutrino mass
once the rate of the $0\nu\beta\beta$ decay have been determined.
It is important to stress, however, that quite generally an 
observation of {\bf any} total
lepton number violating process, not only of the $0\nu\beta\beta$ decay, 
would necessarily imply that neutrinos are massive Majorana fermions.

\section{$2\nu\beta\beta$ decay}

Study of the $2\nu\beta\beta$ decay is an important nuclear physics 
problem by itself. Moreover, evaluation of the $M^{2\nu}$ matrix elements
is an important test for the nuclear theory models that aim at the determination
of the analogous but different quantities for the more fundamental $0\nu$ neutrinoless mode.
So, we begin our discussion with that experimentally more accessible $\beta\beta$
decay mode.

The rate of the $2\nu\beta\beta$ decay was
first estimated by Maria Goeppert-Meyer already in 1937 in her thesis work
suggested by E. Wigner, basically correctly.
Yet, first experimental observation in a laboratory
experiment was achieved only in 1987, fifty years later \cite{Moe}.
(Earlier observations of the $\beta\beta$ decay \cite{Ing50, Kir83, Ber92} were based
on the geochemical method that cannot separate the $2\nu\beta\beta$
and $0\nu\beta\beta$ decay modes. Later laboratory measurements have shown,
as expected, that the geochemical method determines dominantly the  $2\nu\beta\beta$
decay mode.)
 
Note that such delay is not really exceptional in neutrino physics. It took more
than twenty years since the original suggestion of Pauli to show
that neutrinos are real particles in the pioneering experiment by
Raines and Cowan. And it took another almost fifty years since that
time to show that neutrinos are massive fermions.
Why it took so long in the case of the $\beta\beta$ decay?
As pointed out above, the typical half-life
of the $2\nu\beta\beta$ decay is $\sim 10^{19-20}$ years. Yet, its
``signature" is very similar to natural radioactivity, present to some extent
everywhere, and governed by the half-life of $\sim 10^{10}$ years,
or much less for most of  the man-made or cosmogenic radioactivities.
So, background suppression is the main problem to overcome
when one wants to study either of the $\beta\beta$ decay modes.

\begin{figure}[htb]
\centerline{\psfig{file=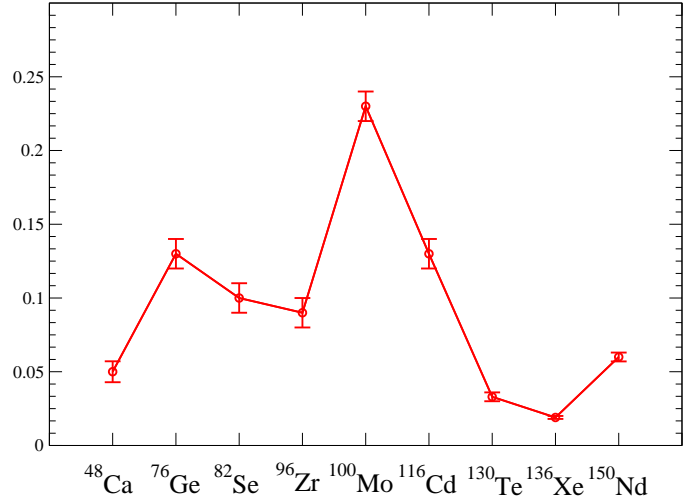,width=8.0cm}}
\caption{ Matrix elements $M^{2\nu}$ in MeV$^{-1}$ based on the
experimental halflife measurements. }
\label{fig_2nu}
\end{figure}

During the last two decades the $2\nu\beta\beta$ decay has been observed
in ``live" laboratory experiments
in many nuclei, often by different groups and using different
methods. That shows not only the ingenuity of the experimentalists who
were able to overcome the background nemesis, but makes it possible
at the same time to extract the corresponding $2\nu$ nuclear matrix element
from the measured decay rate. The resulting nuclear matrix elements $M^{2\nu}$, 
which have the dimension energy$^{-1}$,
are plotted in Fig.\ref{fig_2nu}. Note the pronounced shell dependence; the matrix element
for $^{100}$Mo is almost ten times larger than the ones for $^{130}$Te or $^{136}$Xe. \\

The derivation of the $2\nu\beta\beta$-decay rate formula
is analogous to the treatment of ordinary beta decay. It
begins with the Fermi golden rule for second order weak decay
\begin{equation}
\frac{1}{\tau} = 2\pi \delta(E_0 - \Sigma_f E_f)
\left[ \Sigma_{m,\mu} \frac{\langle f| H_{\mu} | m \rangle \langle m| H^{\mu} | i \rangle}
{E_i - E_m - p_{\nu} - E_e} \right]^2 ~,
\end{equation}
where the sum over $m$ includes all relevant virtual states in the intermediate odd-odd
nucleus and $\mu$ labels the different Dirac structures
of the weak interaction Hamiltonian.

When calculating the $0^+ \rightarrow 0^+ $ transitions,
it is generally a very good approximation
to replace the lepton energies in the denominator by the corresponding average value, i.e.,
$E_e  +  p_{\nu}  \sim E_0 /2$, where the $E_0 = M_i - M_f$ is the total decay
energy including electron masses. The lepton momenta, for both electrons and neutrinos 
are all $q < Q$ and thus
$qR \ll 1$, where R is the nuclear radius. Hence the so-called long wavelength
approximation is valid and the rate formula  separates
into a product of the nuclear and lepton parts, where the lepton part $G^{2\nu}(E_0,Z)$
is just the phase-space integral discussed and tabulated earlier.

The nuclear structure information is contained in the nuclear matrix element;
only the Gamow-Teller $\sigma \tau$ part contributes in the long wavelength
approximation
\begin{equation}
M^{2\nu} = \Sigma_{m,i,k} \frac { \langle 0^+_f | \vec{\sigma}_i \tau^+_i | m \rangle
\langle m | \vec{\sigma}_k \tau^+_k | 0^+_i \rangle}{E_m - (M_i + M_f)/2} ~.
\label{eq_m2nu}
\end{equation}
The individual terms in the eq. (\ref{eq_m2nu}) have a well defined meaning, in particular
for the most relevant ground state to ground state transitions. The terms
$\langle m | \vec{\sigma}_k \tau^+_k | 0^+_i \rangle$ represent the 
amplitudes of the $\beta^-$
strength of the initial nucleus and can be explored in the nucleon charge exchange
reactions such as $(p,n)$ and $(^3He,t)$. On the other hand the terms
$\langle 0^+_f | \vec{\sigma}_i \tau^+_i | m \rangle$ represent the $\beta^+$
strength in the final nucleus and can be explored in the opposite nucleon charge exchange
reactions such as $(n,p)$ and $(d, ^2He)$. In this way one can (up to the sign)
explore the contribution of several low lying states to the $M^{2\nu}$ matrix element.
(See the article by H. Ejiri and D. Frekers in this issue.)

In Eq.(\ref{eq_m2nu}) the energy denominators signify that the low-lying intermediate
$I^{\pi} = 1^+$ states contribute significantly more that the high-lying states.
Nevertheless, it is useful to analyze a related expression obtained in the closure 
approximation:
\begin{equation}
M^{2\nu} = \Sigma_{m,i,k} \frac { \langle 0^+_f | \vec{\sigma}_i \tau^+_i | m \rangle
\langle m | \vec{\sigma}_k \tau^+_k | 0^+_i \rangle}{E_m - (M_i + M_f)/2} 
\equiv \frac{  \langle 0^+_f | \Sigma_{i,k}  \vec{\sigma}_i \tau^+_i
\cdot \vec{\sigma}_k \tau^+_k  | 0^+_i \rangle}{\Delta \bar{E}} ~,
\label{eq_clo}
\end{equation}
where $\Delta \bar{E}$ is the average energy denominator defined by the above equation.
The numerator of Eq.(\ref{eq_clo}) is the definition of the closure $2\nu$ nuclear
matrix element. By itself, the $2\nu$ closure  matrix element 
does not have a direct physics interpretation.

For the two-body operator (i.e. in closure) 
in the $2\nu$ matrix element one can derive an approximate
sum rule, analogous to the famous Ikeda sum rule of the GT operator \cite{Magda}.
Remember that for the beta strengths 
$S^{\beta\mp} = \Sigma \vec{\sigma}_i \tau_i^{\pm}$ the Ikeda sum rule is obeyed
independently of the nuclear structure as long as nuclei are assumed to be made
of nucleons only
\begin{equation}
S^{\beta-} - S^{\beta+} = 3(N-Z)~.
\label{eq:Ikeda}
\end{equation}
In nuclei with neutron excess, $N > Z$ the $\beta^+$ strength is much smaller than
the $\beta^-$ strength, which is concentrated in the giant GT resonance.

In analogy, the strength corresponding to the double beta operator 
$\Sigma_{i,k} \vec{\sigma}_i \tau_i^+  \vec{\sigma}_k \tau_k^+$ will be
concentrated mostly in the ``double GT resonance'' and its strength will
be approximately equal to the sum rule $6(N-Z)(N-Z+1)$ \cite{Magda}. Assuming further that
the average energy denominator $\Delta \bar{E}$ in Eq. (\ref{eq_clo}) is $O(1)$ MeV,
we see that $2\nu\beta\beta$ closure matrix element, connecting the ground states
of the initial and final nuclei (see Fig. \ref{fig_2nu}) is heavily suppressed,
representing only a very small fraction of the corresponding sum rule.
From that it follows that an accurate evaluation of the matrix elements $M^{2\nu}$
and $M^{2\nu}_{cl}$ is going to be quite difficult, since these quantities are
so small in their natural units.

\begin{figure}[htb]
\centerline{\psfig{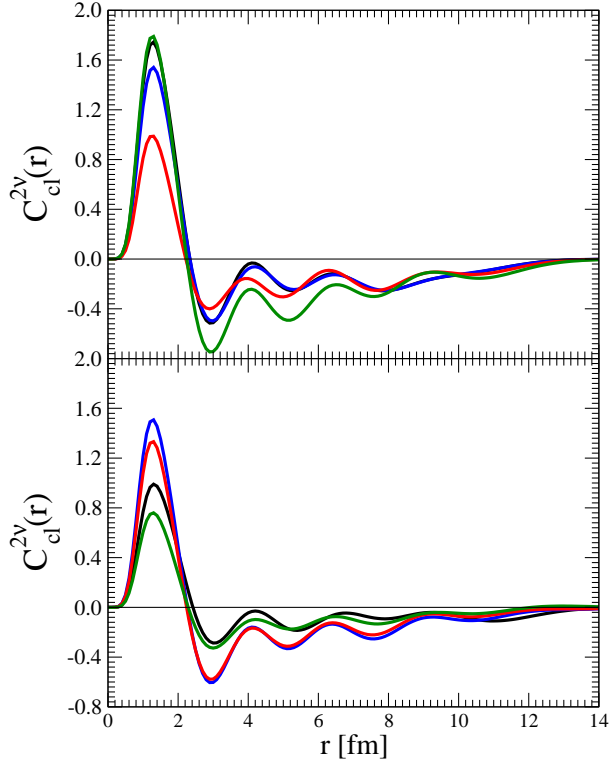}}
\caption{ Functions $C^{2\nu}_{cl}(r)$ for several nuclei using
the QRPA method. In the upper panel are $^{76}$Ge (black), $^{82}$Se (blue),
$^{96}$Zr (red), and $^{100}$Mo (green). In the lower
panel are $^{116}$Cd (black), $^{128}$Te (blue), ${130}$Te (red), and $^{136}$Xe (green).}
\label{fig_C2nu(r)}
\end{figure}

To shed more light on the problem, consider the dependence of the $2\nu\beta\beta$
matrix elements on the distance between the two neutrons that are transformed into
two protons. These two neutrons are not correlated, but nevertheless they are both
bound in the corresponding nuclei, and decay together. To characterize the radial
dependence lets introduce the function
\begin{eqnarray}
C^{2\nu}_{cl}(r) =  \langle f | \Sigma_{lk} \vec{\sigma}_l \cdot  \vec{\sigma}_k
\delta(r - r_{lk}) \tau_l^+ \tau_k^+ | i \rangle ~, \nonumber \\
M^{2\nu}_{cl} = \int_0^{\infty} C^{2\nu}_{cl}(r) dr ~.
\label{eq:2nuclp}
\end{eqnarray}
This definition is in analogy to the related function $C^{0\nu}(r)$ first introduced in
Ref. \cite{EV04}, and discussed in detail later. Note that while the matrix elements 
$M^{2\nu}$ and $M^{2\nu}_{cl}$ get contributions only from the $1^+$
intermediate states, the function $C^{2\nu}_{cl}$ gets contributions
from all intermediate multipoles.
This is the consequence of the $\delta$ function in the definition of $C^{2\nu}_{cl}(r)$.
When expanded, all multipoles contribute.
Naturally, when integrated over $r$ only the contributions
from the $1^+$ are nonvanishing. Examples of the function $C^{2\nu}_{cl}(r)$ are
shown in Fig. \ref{fig_C2nu(r)}. (Similar figure appears in Ref. \cite{5th}.)

\vspace{1cm}

One can see that in all the cases shown in Fig. \ref{fig_C2nu(r)} the function $C^{2\nu}_{cl}(r)$
consists of a positive peak at $r \sim 1$ fm and a negative tail starting at $r \sim 2$ fm. It turns out
that the areas under the positive peak and the negative tail are nearly equal, resulting in
considerable uncertainty in $M^{2\nu}_{cl}$.

Theoretical evaluation of the matrix elements $M^{2\nu}$ and $M^{2\nu}_{cl}$, respectively 
application of the corresponding rate equation (\ref{eq:rate}), involves another problem,
namely whether one should, or should not, apply the experience with the ordinary $\beta$ decay
and use the concept of quenching of the GT strength \cite{quen}. We shall discuss that issue
in more detail later.

\section{Operator of the $0\nu\beta\beta$ decay}

The $0\nu$ decay rate associated with the nonvanishing value of $m_{\nu}$ is of the
general form
\begin{equation}
\omega_{0\nu}  =  2 \pi \Sigma_{spin} | R_{0\nu} |^2
\delta (E_{e1}  +  E_{e2}  +  E_f - M_i ) d^3 p_{e1} d^3 p_{e2} ~,
\end{equation}
where $E_f $ is the energy of the final nucleus and $R_{0\nu}$ is the
transition amplitude including both the lepton and nuclear parts.

After substitution for the neutrino propagator
and integration over the virtual neutrino momentum,
the lepton amplitude acquires the form
\begin{equation}
-i \delta_{jk} \int \frac{d^4 q}{ (2 \pi )^4 }
\frac{e^{ -iq(x-y) }}{ q^2 - m_j^2 }
\bar{e} (x) \gamma_{\rho} \frac{1}{2} (1 - \gamma_5 )
( q^{\mu} \gamma_{\mu} + m_j ) \frac{1}{2} (1 - \gamma_5 )
\gamma_{\sigma}  e^C (y) ~.
\end{equation}

From the commutation properties of the gamma matrices it then follows that
the decay amplitude for purely left-handed lepton currents is proportional
to the neutrino Majorana mass $m_j$. Integration over the virtual neutrino 
energy leads to the replacement of the propagator
$(q^2 - m_j^2 )^{-1}$ by the
residue $\pi/ \omega_ j$  with $ \omega_ j = ( \vec{q}^{~2} + m_j^2 )^{1/2}$.

Finally, the integration over the space part $d \vec{q}$
leads to an expression for the "neutrino
potential" that appears in the corresponding nuclear transition operator,
\begin{equation}
H(r,E_m )  =  \frac{R}{ 2 \pi^2}  \int  \frac{ d \vec{q}}{ \omega}
\frac {1}{ \omega + A_m }  e^{ i \vec{q} \cdot \vec{r}}  =
\frac{2R}{ \pi r}  \int_0^ {\infty} dq \frac{ q \sin (qr) }{ \omega( \omega + A_m ) }
= \frac{2R}{\pi} \int_0^ {\infty} dq \frac{j_0(qr)q}{q + A_m}  ~.
\label{eq_nupot}
\end{equation}
where the nuclear radius $R = 1.2A^{1/3}$ fm was added as an auxiliary factor
so that $H$ becomes dimensionless. A corresponding $1/R^2$ compensates
for this auxiliary quantity in the phase space formula.
The weak dependence on the excitation energy
of the virtual intermediate odd-odd nucleus appears in 
$A_m = E_m -  E_i  +  E_e \equiv E_m - (M_i - M_f)/2$.

The momentum of the virtual neutrino is
determined by the uncertainty relation
$q \sim 1/r$, where $r \le R$ is a typical spacing between two nucleons.
We will show later that in fact the relevant values of $r$ are only $r \le$ 2-3 fm,
so that the momentum transfer $q \sim$ 100-200 MeV. For the light neutrinos
the neutrino mass $m_j$ can then be safely neglected in the potential $H(r)$.
(Obviously, for heavy neutrinos, with masses $M_j \gg$ 1 GeV a different procedure
is necessary.) Also, given the large value of $q$ the dependence on the
difference of nuclear energies $E_m - E_i$ is expected to be rather weak
and the summation of the intermediate states can be performed in closure
for convenience. This approximation $H(r,E_m) \simeq H(r, \bar{E})$ is, 
in fact, typically used in the evaluation of the $M^{0\nu}$,
where $\bar{E} \sim 5-10$MeV is the characteristic nuclear excitation energy.

\begin{figure}[htb]
\centerline{\psfig{file=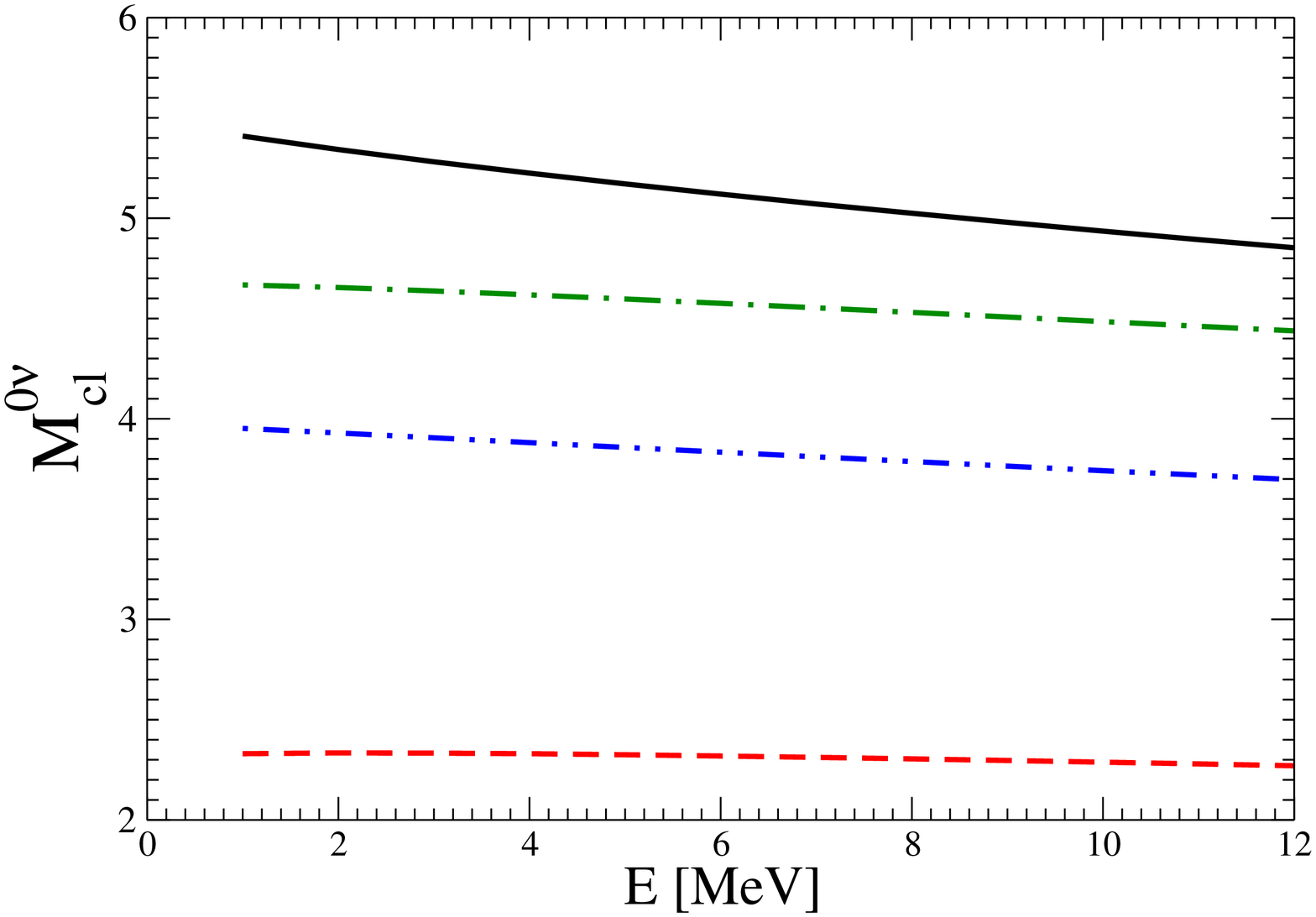,width=9.0cm}}
\caption{
Matrix elements $M^{0\nu}$ for several nuclei evaluated within QRPA
in the closure approximation as a function of the assumed average excitation
energy.
The values of $M^{0\nu}$ obtained without the closure
approximation are 5.24 ($^{76}$Ge, full black line), 2.62 ($^{96}$Zr, dashed red line ), 
4.99 ($^{100}$Mo, dot-dashed green line ), and 4.07 ($^{130}$Te, double dot-dashed blue line).}
\label{fig:cl}
\end{figure}

It is worthwhile to test the validity of this approximation. Such test can be conveniently
performed within the QRPA, where the sum over the intermediate states can be easily explicitly carried out. 
In this context one can ask two questions: How good is the closure approximation? And
what is the value of the corresponding average energy?  In Fig. \ref{fig:cl} we illustrate the
 answers to these questions (see also \cite{5th}). The QRPA matrix elements evaluated by explicitly summing over the 
virtual intermediate states quoted in the caption can be compared with the
curves obtained by replacing all intermediate energies with a constant $\bar{E}$,   
which is varied there between 0 and 12 MeV.
One can see, first of all, that the $M^{0\nu}$ changes modestly, by less than 10\% when $\bar{E}$
is varied as expected given the relative sizes of $q$ and  $\bar{E}$ and, 
at the same time, that the exact results are quite close, but somewhat larger,
than the closure ones. Thus, employing
the closure approximation is appropriate for the evaluation of $M^{0\nu}$
even though it apparently slightly underestimates the $M^{0\nu}$ values. \\ \\

The neutrino potential in the Eq. (\ref{eq_nupot}) was defined assuming that the nucleons
are point particles. That is not true, however, and thus it is necessary to include 
a corresponding correction in the definition  of $H(r,\bar{E})$. It is customary to approximate this
correction in the form of the dipole type form factor
\begin{equation}
f_{FNS} = \frac{1}{\left(1 + \frac{q^2}{M_A^2}\right)^2} ~,
\label{eq:fns}
\end{equation}
with $M_A = 1.09$ GeV. Varying $M_A$ in the interval 1.0-1.2 GeV makes little difference.

In addition, while the weak current of quarks has the simple $V-A$ structure, the weak 
nucleon current contains additional terms, since nucleons are complicated composite objects, 
\begin{equation}
J^{\mu +} = \Psi \tau^+ \left[ g_V(q^2)\gamma^{\mu} - ig_M(q^2)\frac{\sigma^{\mu \nu}}{2 m_p} q_{\nu}
-g_A(q^2)\gamma^{\mu}\gamma_5 + g_P(q^2)q^{\mu}\gamma_5 \right] \Psi ~,
\end{equation}
representing the vector, weak magnetism, axial vector and induced pseudoscalar. 
In the zero momentum transfer limit the values of the corresponding form factors are well known.
The induced pseudoscalar form factor is usually evaluated using the partially conserved axial-vector
current hypothesis
\begin{equation}
g_P(q^2) = \frac{g_A(q^2) \times 2m_p}{q^2 + m_{\pi}^2} ~,
\end{equation}
and the weak magnetism form factor is simply proportional to $g_V(q^2), g_M(q^2) = (\mu_p - \mu_n)g_V(q^2)$.

Taking these ``higher order terms'' into account and using the nonrelativistic limit
for the nucleons, one arrives at the corresponding correction term $g_{HOT}(q^2)$ in the
main, GT part of the neutrino potential (for details see Ref. \cite{Sim99})
\begin{equation}
g_{HOT}^{GT}(q^2) =  1 - \frac{2}{3}\frac{\vec{q}^2}{\vec{q}^2 + m_{\pi}^2} + \frac{1}{3}
\left( \frac{\vec{q}^2}{\vec{q}^2 + m_{\pi}^2} \right)^2 + 
\frac{2}{3} \left(\frac{g_V(q^2)}{g_A(q)} \right)^2 \frac{(\mu_p - \mu_n)^2 \vec{q}^2}{4 m_p^2}  ~.
\end{equation} 

Finally, the neutrino potential $H_{GT}(r,\bar{E})$ governing the Gamow-Teller part of the matrix element
$M^{0\nu}$ with these correction factors included is of the form
\begin{equation}
 H_{GT}(r,\overline{E}_{0\nu}) =  
\frac{2R}{\pi} \int_0^{\infty} j_0(q r)\frac{q}{q + \overline{E}_{0\nu}}
f^2_{FNS}(q^2)g_{HOT}^{GT}(q^2)dq ~,
\label{eq:pot}
\end{equation}

\begin{figure}[htb]
\centerline{\psfig{file=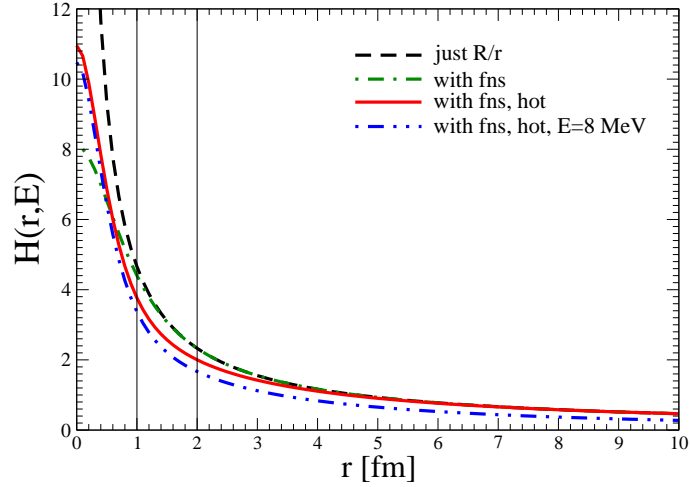,width=9.0cm}}
\caption{The potential $H_{GT} (r, \bar{E})$. Different approximate forms, as well as the exact one,
are shown.}
\label{fig:pot}
\end{figure}

When the finite nucleon size,
higher order terms are neglected, and $\bar{E}_{0\nu} = 0$ is assumed, the potential has
Coulomb-like shape $R/r$. The full potential, Eq. (\ref{eq:pot}), however, is finite at $r=0$,
$H(r \rightarrow 0, \bar{E}_{0\nu}=0) = 5M_A R/16$.  Including the higher order currents
and finite $\bar{E}$ in Eq. (\ref{eq:pot}) increases the value of  $H(r=0)$ by $\sim$30\%.
The shape of the neutrino potential is shown in Fig. \ref{fig:pot}. \\ \\

In the full $0\nu\beta\beta$ operator there are several potentials, each with its own spin structure 
\begin{equation}
O^{0\nu} = \Sigma_{i,k} \tau_i^+ \tau_k^+ \left(-H_F(r,\overline{E}_{0\nu})\frac{g_V^2}{g_A^2}  
+ H_{GT}(r,\overline{E}_{0\nu}) \vec{\sigma}_i \cdot \vec{\sigma}_k +  
H_{T}(r,\overline{E}_{0\nu})S_{ik} \right) ~,
\end{equation}
where in the momentum representation $S_{ik} = 3 \vec{\sigma}_i \cdot \vec{q} 
\vec{\sigma}_k \cdot \vec{q} -  \vec{\sigma}_i \cdot \vec{\sigma}_k$ and
$g_{HOT}^{F}(q^2) = 1$, $g_{HOT}^{T}(q^2) = \frac{2}{3} \frac{\vec{q}^2}{\vec{q}^2 + m_{\pi}^2} -
 \frac{1}{3} \left( \frac{\vec{q}^2}{\vec{q}^2 + m_{\pi}^2} \right)^2$. Also, in $f_{FNS}$ for the
Fermi part one needs to replace $M_A$ by $M_V \simeq 0.85$ GeV, and in the tensor part
$g_{HOT}^{T}(q^2)$
the spherical Bessel function $j_0(qr)$ must be replaced by $j_2(qr)$.

To finish this section, lets briefly mention the analysis of the $0\nu\beta\beta$ decay
assuming the existence of the right-handed weak currents. Such phenomenological approach
is often quoted in the literature, even though it is not at all clear that there exist
a corresponding realistic particle physics model giving the $0\nu\beta\beta$-decay rates
competitive with the standard light left-handed Majorana neutrino exchange, like in
Eq. (\ref{eq:rate}). The assumed hamiltonian is
\begin{equation}
H_W = \frac{G_F \cos\theta_C}{\sqrt{2}} \Sigma_{i=1}^{2n} \left[ j_{Li\mu} \bar{J}_{Li}^{\mu \dag}
+ j_{Ri\mu} \bar{J}_{Ri}^{\mu \dag} \right] ~,
\end{equation}
where the lepton currents are 
\begin{equation}
j_{Li}^{\mu} = \bar{e}(1 - \gamma_5)N_{iL} ~;~~~ j_{Ri}^{\mu} = \bar{e}(1 + \gamma_5)N_{iR} ~,
\end{equation}
with the hypothetical heavy neutrinos $N_{iL(R)}$ with $n$ mass eigenstates.
The nuclear currents are
\begin{equation}
\bar{J}_{Li}^{\mu \dag} = U_{ei} J_L^{\mu \dag} ~;
~~~ \bar{J}_{Ri}^{\mu \dag} = V_{ei} [\lambda J_R^{\mu \dag} + \eta J_L^{\mu \dag} ] ~,
\end{equation}
where $U_{ei}$ is the standard light neutrino mixing matrix and $V_{ei}$ is its analog
for the heavy neutrinos. Parameters $\lambda$ and $\eta$ characterize the strength 
of the right-right and right-left interactions.
The formulae for the corresponding $0\nu\beta\beta$-decay rate have been evaluated in many papers, here I
quote just two of them \cite{Doi,Tomoda}. In those references one can find the complete expressions
for all necessary neutrino potentials and the corresponding spin and $\vec{q}$ dependencies.

\section{Decays mediated by the heavy particle exchange}
\label{sect:heavy}

As stressed above, when heavy particles of any kind mediate the $0\nu\beta\beta$ decay, we are dealing with
a six fermion vertex, representing extremely short range operator. Vergados \cite{Ver81} was presumably
the first author to describe how the issue of suppression, due to the short range nucleon-nucleon repulsion,
can be overcome.    

\begin{figure}[htb]
\centerline{\psfig{file=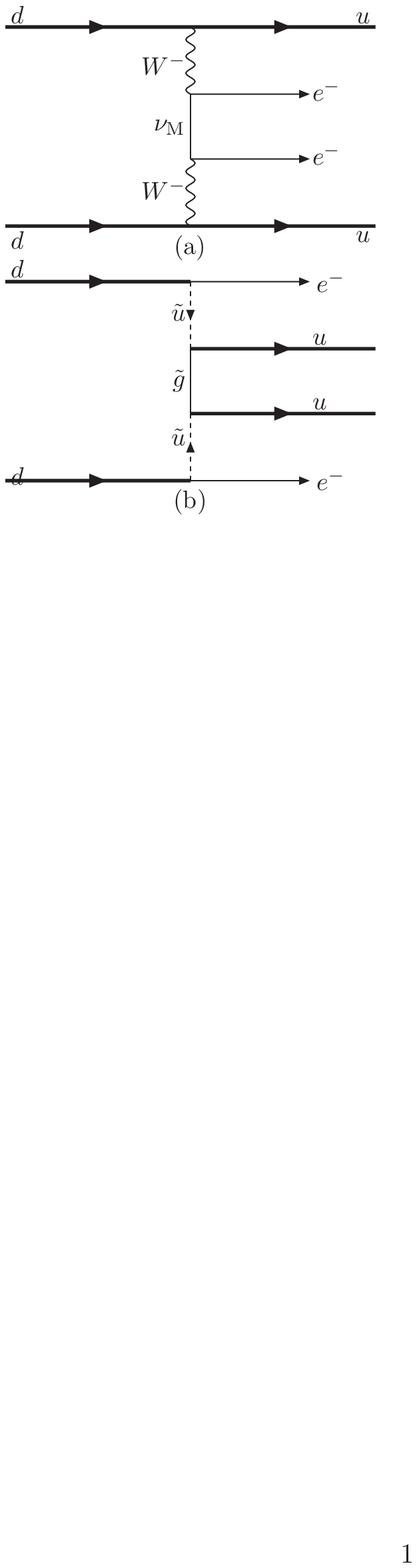,width=6.0cm}}
\caption{Schematic graph indicating the $0\nu\beta\beta$ decay mediated by the heavy neutrino exchange $\nu_M$
in panel (a). In panel (b) the possible $0\nu\beta\beta$ decay through the exchange of other heavy particles,
in this case two squarks and a gluino in RPV SUSY.}
\label{fig:heavy}
\end{figure}


If one could treat nucleons as pointlike particles and  assume that
the heavy neutrino mass will be
substantially larger than the corresponding momentum transfer $q$ (say, $m_{heavy} \gg$ 1 GeV) 
the neutrino potential would contain exp(-$mr$) where $m$ = min$(m_{heavy}, M_W)$.
The corresponding nuclear matrix element would be heavily suppressed due to the extremely short range
of this potential.

However, nucleons are not pointlike particles. Instead, their finite size could be parametrized
through the dipole type form factor $f_{FNS}$ in Eq.(\ref{eq:fns}). In that case the 
neutrino potential will be of the form
\begin{equation}
H_{heavy\nu}(r, M_A) = \frac{4\pi R}{M_A^2} \int \frac{d \vec{q}}{(2\pi)^3} \left( 
\frac{M_A^2}{M_A^2 + \vec{q}^2} \right)^4 = \frac{M_A R}{48} e^{-M_A r} 
\left[ 1 + M_A r + \frac{1}{3} (M_a r)^2 \right]
\end{equation}
Such potential will have the range $1/M_A$ and will be much less affected by the short range correlations
(see e.g. Ref. \cite{Ver81,Tomoda}). The disadvantage of the form factor modeling is that the error introduced
by such approximation is very difficult to estimate.

In general the heavy particle exchange is characterized by some energy scale $\Lambda_{\beta\beta}$
(see Eq. (\ref{eq_estimate}), that is  much larger
than any hadronic scale $\Lambda_H \sim$ 1 GeV that would enter the problem. In that case one could,
instead of the form factor approach indicated above, use the prescriptions
of effective field theory and classify the contributions in powers of small quantities $q/\Lambda_H$, 
$q/\Lambda_{\beta\beta}$ and $\Lambda_H/\Lambda_{\beta\beta}$ as in Ref. \cite{garybb}.

\begin{figure}[htb]
\centerline{\psfig{file=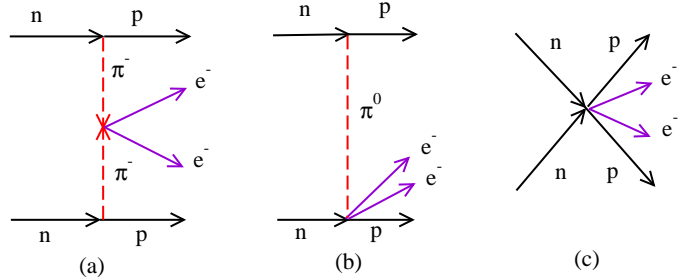,width=9.0cm}}
\caption{Diagrams that contribute to $0\nu\beta\beta$ decay at tree level.
Panel (a) represents $\pi\pi ee$ contribution, panel (b) $NN\pi ee$ contribution 
and (c) $NNNNee$ contribution.}
\label{fig:eft}
\end{figure}

The hadronic vertices appearing in the corresponding Lagrangian will be of the type $NNNNee$, $NN \pi ee$ and
$\pi\pi ee$ as illustrated in Fig. \ref{fig:eft}. They stem from quark-lepton operators 
having different transformation properties
under parity and chiral SU(2). As such, they will contribute to 
different orders in the $q/\Lambda_H$ expansion.
The vertices involving pions are longer range. They have been analyzed in the form factor 
approach in Ref. \cite{Ver82},
but the EFT allows more systematic approach because of the separation of scales 
$q < \Lambda_H \ll \Lambda_{\beta\beta}$.
It was noted already in Ref. \cite{Fae97} that the nuclear matrix elements associated 
with the long range pionic effects within the RPV SUSY scenarios can be dominant. But that
is, in fact, a more general result. The pionic effects can be substantially larger
than those obtained using the conventional form factor model for the short-range
$NNNNee$ process.    
 

\vspace{4cm}

The lepton number violating vertices in Fig.\ref{fig:eft} represent nonstandard model operators,
and can be characterized by some parameters $K_{\pi\pi}$, $K_{NN\pi}$ and $K_{NNNN}$. Counting
powers of $q/\Lambda_H$, for $\Lambda_H = 4\pi f_{\pi}, ~f_{\pi} \simeq$ 92 MeV, we come
to the conclusion that 
\begin{equation}\label{pcounting}
\rm{Fig.~\ref{fig:eft}a}\sim K_{\pi\pi} p^{-2},
~~~ \rm{Fig.~\ref{fig:eft}b,} \sim K_{NN\pi}p^{-1},
~~~ \rm{Fig.~\ref{fig:eft}c} \sim K_{NNNN}p^0,
\end{equation}
Thus, the long range $0\nu\beta\beta$-decay operators $\pi\pi ee$ and to a lesser
degree $NN\pi ee$ are enhanced, relative to the
short range operator $NNNNee$ in Fig.\ref{fig:eft}(c). 

Clearly, the most important operator is the one corresponding to the $\pi\pi ee$ vertex.
It was shown in Ref. \cite{garybb} that the corresponding Lagrangian can be expressed
as
\begin{equation}\label{eq:ppee}
{\cal{L}}_{(0)}^{\pi\pi ee}
 = \frac{G_{\rm{F}}^2\Lambda_{\rm{H}}^2
  f_\pi^2}{\Lambda_{\beta\beta}}\left\{
\pi^-\pi^-\bar{e}(\beta_1 + \beta_2\gamma^5)e^c +
\pi^+\pi^+\bar{e}^c(\beta_1 - \beta_2 \gamma^5)e
\right\}~,
\end{equation}
where $\beta_{1,2}$ are dimensionless parameters that need be evaluated in any
concrete particle physics model. 

Transforming the above Lagrangian into the hadron-lepton system, taking the
nonrelativistic limit and Fourier transforming to coordinate space yield an
expression for the transition matrix element 

\begin{eqnarray}\label{eq:heavym}
{\cal{M}}_0 &=& <\Psi_{A,Z+2}|
\sum\limits_{ij}\frac{R}{r_{ij}}
[ F_1\!(x_{ij}) \vec{\sigma}_i \cdot \vec{\sigma}_j
+ F_2\!(x_{ij}) S_{ij} ] \tau^+_i\tau^+_j
 | \Psi_{A,Z} >,
\\ \nonumber
S_{ij} &=& 3\vec{\sigma}_i \cdot \hat{\vec{r}}_{ij}
\vec{\sigma}_j \cdot \hat{\vec{r}}_{ij} - \vec{\sigma}_i \cdot
\vec{\sigma}_j~.
\end{eqnarray}
Here $r_{ij}$ is the distance between the $i$th and $j$th neutrons, $\hat{\vec{r}} = \vec{r}/r$ 
and $x_{ij}=r_{ij}m_{\pi}$.
The form-factors $\rm{F}_1$ and $\rm{F}_2$ were first introduced
in Ref.~\cite{Ver82}
\begin{equation}
\rm{F}_1(x)=(x-2)\rm{e}^{-x},~~~~~~~
\rm{F}_2(x)=(x+1)\rm{e}^{-x}~.
\end{equation}

The $0\nu\beta\beta$ halflife is, as expected, inversely proportional to the square of the scale
$\Lambda_{\beta\beta}$ and depends on the empirical parameters $\beta_{1,2}$, 
\begin{eqnarray}\label{eq:heavyt}
\frac{1}{T_{1/2}} &=&
\frac{\hbar c^2}{144\pi^5\ln\!2}\frac{g_{\rm{A}}^4}{R^2}
  \frac{\Lambda_{H}^4 G_F^4}{\Lambda_{\beta\beta}^{2}}
  \int_{m_e}^{E_{\beta\beta} - m_e}\rm{d}\!E_1
  F\!(Z+2,E_1)F\!(Z+2,E_2) ~~~~
\nonumber \\
  & & ~~~~~~~~~~~~~~~~~~\frac{1}{2} [(\beta_1^2+\beta^2_2)p_1E_1p_2E_2
  -   (\beta_1^2-\beta^2_2)p_1p_2m_e^2] |{\cal{M}}_0|^2,
\end{eqnarray}

As mentioned earlier, concrete application of the pion exchange mechanism
in Ref. \cite{Fae97} suggests the dominance of the $\pi\pi ee$ over the
short range nucleon only vertex by a factor of 10 - 30 at the level of the
nuclear matrix element. This is, to some extent, so far largely unexplored
issue. The application of the form factor method, obviously insufficient,
is more or less a norm even today (see e.g. the recent paper \cite{Rath12}).

\section{Short range correlations}
\label{sect:short}

\begin{figure}[htb]
\centerline{\psfig{file=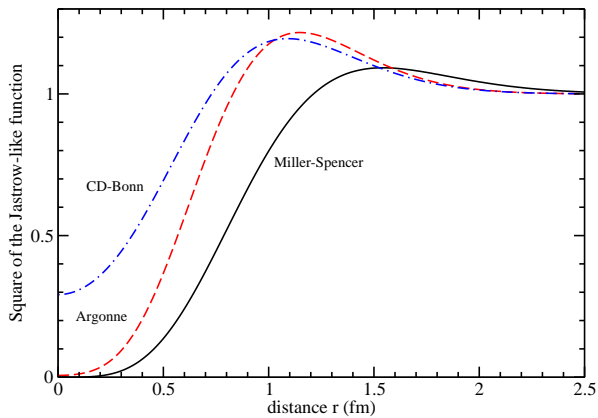,width=7.0cm}}
\caption{Plots of the squares of the Jastrow-like correlation functions described in the text.}
\label{fig:short}
\end{figure}

When evaluating the $0\nu\beta\beta$ nuclear matrix element in the
closure approximation  it is necessary, ultimately,
to consider the matrix element of a two-body transition operator which connects
a state with two neutrons in a specific mean field single particle states with two protons
again in some specific final states. 
These states are angular momentum and possibly
isospin coupled single particle states. In essentially all cases they do not explicitly
include the effect of the short range nucleon-nucleon repulsion.

Until recently, it was customary to simulate the effect of such short range correlations
by including in the operator (or equivalently in the two body wave-function) a
phenomenological Jastrow-like function
\begin{equation}
f_{Jastrow}(r) = 1 - c e^{-ar^2}(1 - br^2)~.
\label{eq:Jas}
\end{equation}
In other words,
\begin{equation}
M^{0\nu} = \Sigma_{k<m} M_{km} ~~~ M_{km} \rightarrow f(r_{km})M_{km} f(r_{km})
\end{equation}
In early works the parameters $a,b,c$, determined for a different purpose in Ref. \cite{Miller},
were used ($c = 1.0, a =$ 1.1 fm$^{-2}$ and $b = $0.68 fm$^{-2}$). 
This is the so called Miller-Spencer parametrization. Since in the evaluation of the
nuclear matrix element such function appears twice, we plot in Fig. \ref{fig:short} its square.
Application of this prescription resulted typically in the reduction of the corresponding
 $0\nu\beta\beta$ nuclear matrix element by $\sim$30\% when compared to the result when
$f_{Jastrow}(r) = 1$ is used.

Recent works have questioned the adequacy of this prescription.
One of the approaches proposed relatively
recently in \cite{UCOMa} is based on the Unitary Correlation
Operator Method (UCOM), see Ref.\cite{UCOM}. This approach
describes not only short-range, but also central and tensor correlations
explicitly by means of a unitary transformation.
Applied to a realistic
NN interaction, the method produces a correlated interaction,
which can be used as a universal effective interaction. In the case
of the $0\nu\beta\beta$-decay calculation the
correlated two-nucleon wave function was taken as 
\begin{equation}
| {{\Psi}}_{\mathcal J} \rangle =
C_r | {{\Psi}}_{\mathcal J} \rangle.
\end{equation}
Here, $C_r$ is the unitary correlation operator
describing the short-range correlations. The explicit
form of $C_r$ is given in \cite{UCOM}.
The UCOM-corrected nuclear matrix elements are
significantly less reduced when compared with
those calculated with Jastrow SRC \cite{anatomy,UCOMa}.
In practise, the UCOM prescription requires that the operator $r_{ab}$
be replaced by a shifted $R_+(r_{ab})$. As pointed out in Ref. \cite{Jon11},
care must be taken to apply the prescription consistently, not just simply
by replacing the neutrino potential $H(r) \rightarrow H(R_+(r))$.  

In Refs. \cite{Jon09,Sim09} the effect of short range correlations was computed
within well defined Brueckner-based approximation scheme. In particular,
in \cite{Sim09} the coupled cluster method was used, based on the realistic $NN$ interactions,
CD-Bonn and Argonne V18. In a good approximation the effect of correlations
can be again approximated by the Jastrow-like functions (see Eq. (\ref{eq:Jas}))
with parameters fitted in Ref. \cite{Sim09}, $a=1.59$ fm$^{-2}$, $b=1.45$ fm$^{-2}$
and c=0.92 for the case of the Argonne potential and  $a=1.52$ fm$^{-2}$, $b=1.88$ fm$^{-2}$
and c = 0.46 for the case of the Argonne potential. Both of these functions
are displayed in Fig. \ref{fig:short}. It is obvious that the application of these
functions leads to the much less reduction of the $M^{0\nu}$, in agreement
with the UCOM method and with the conclusion of Ref. \cite{Jon09} than the Miller-Spencer
approach used previously. 
In fact, provided the nucleon finite size form factor are properly taken into
account, the application of the UCOM procedure or the prescripition of Ref. \cite{Sim09}
leads to the  $M^{0\nu}$ values essentially unchanged in comparison to the
omission of the short range correlation correction altogether. Thus, the  $M^{0\nu}$
are about 30\% larger than after the application of the traditional Miller-Spencer
parametrization.
There is a general consensus that the matter of treatment
of the short range correlation is satisfactorily resolved at the present time.

\section{$0\nu\beta\beta$ nuclear matrix elements: Basics}

In this section the basic issues involved in the numerical evaluation
of the $0\nu\beta\beta$ nuclear matrix elements will be described. 
We concentrate here on the standard
light Majorana neutrino exchange mechanism, so that the relation between
the decay rate and the nuclear matrix elements is governed by Eq. (\ref{eq:rate}). 
First, there are few reasonable
assumptions common to all (or essentially all) methods:
\begin{itemize}
\item In the $0^+ \rightarrow 0^+$ transitions the outgoing leptons are in
the $s_{1/2}$ (or more precisely Dirac's $\kappa = -1$) state. That assumption
is used in the
calculation of the phase space factor.
\item The hadronic currents are treated in the nonrelativistic impulse
approximation.
\item In most, but not all,  applications the closure approximation is used, i.e.
the matrix element of a two-body operator is considered (see Fig. \ref{fig:cl}).
\end{itemize}

Since the decay involves the transformation of two bound neutrons into two bound
protons, it is necessary, first of all, to choose the proper mean field in which the
nucleons are bound. That field could be spherical, but it can be also deformed.
In the case of deformed mean field the corresponding intrinsic states have no
definite angular momentum; projection into states with definite value of $J^{\pi}$
is sometimes performed in that case.

Once the mean field has been specified, the set of single particle states $|\psi_i \rangle$
is obtained. The  next step is to decide which of these
single particle states are fully occupied or totally empty in both the initial
and final nuclei. Such states, obviously, cannot participate in the decay;
the fully occupied states form the inert core. The remaining states, i.e. 
the single particle states that are
partially occupied or where the occupancies in the initial and final nuclei
are different, form the valence space. To obtain the resulting many-particle states
one has to take into account the residual interaction among the nucleons constrained
in the valence space. 

While such division into occupied, valence, and empty states appears to be reasonable
and well defined,
in practice real nuclei have diffuse Fermi levels and it is not a priori clear
how to properly decide to which category a given single particle state belongs.
Moreover, the effect of the core and empty states should be included, in principle,
in the renormalization of the nuclear hamiltonian and in the definition of the
effective operators. See Ref. \cite{Jon09} and the contribution of J. Engel
to this focus issue for explanation and references; there
are not many explicit applications of this procedure to the problem at hand, so far.
Ideally, moreover, the mean field should be determined selfconsistently, so that 
the same hamiltonian, or nucleon-nucleon interaction, is used in its definition
and in the treatment of the interaction of valence nucleons. Note that for
ease of computation the single particle wave functions are typically 
taken to be the
eigenfunctions of the harmonic oscillator potential, or superpositions
of such functions.

Variety of methods has been used for evaluation of the $0\nu\beta\beta$
nuclear matrix elements. They differ in their choice of the valence space,
interaction hamiltonian and the ways the corresponding equations of motion 
are solved. Some of these methods are described in detail in separate
contributions to this focus issue. Here I wish to characterize each 
method very briefly and show its typical output. Let me stress that
an exact {\it ``ab initio}'', i.e. without approximations, calculation of $M^{0\nu}$ for the
candidate nuclei is impossible at the present time.  

The nuclear shell model (NSM) is, in principle, the method that seems to be
well suited for this task. In it, the valence space consists of just few
single particle states near the Fermi level (usually one main shell). 
With interaction that is based on the realistic nucleon-nucleon force,
but renormalized slightly to describe better masses, energies and transitions
in real nuclei, all possible configurations of the valence nucleons
are included in the calculation. 
The resulting states have not only the correct number of protons and neutrons, 
but also all relevant quantum numbers (angular momentum, isospin, etc). For most nuclei
of interest ($^{48}$Ca is an exception) the valence space, however, does not include
enough states to fulfill the Ikeda sum rule (see Eq.(\ref{eq:Ikeda})), hence 
full description of the $\beta$ strength functions $S^{\beta \pm}$ is not possible.
However, NSM is well tested, since it is capable to describe quite well the spectroscopy 
of low lying states in both initial and final nuclei. In the following figures
\ref{fig:nme},\ref{fig:plotm} and \ref{fig:plott} the NSM results are denoted
by the blue squares.  

The $2\nu\beta\beta$ decay matrix elements $M^{2\nu}$ for several nuclei in Table \ref{tab:1}
are reasonably well described in the NSM, see Ref. \cite{Cau12} ($^{100}$Mo being a notable
exception). However, to achieve this task, it was necessary to apply quenching
factors that, for nuclei heavier than $^{48}$Ca, are considerably smaller than
in the lighter nuclei where the valence space contains the full oscillator
shell. Note that no quenching is applied to the results shown in Figures
\ref{fig:nme},\ref{fig:plotm} and \ref{fig:plott}. I will describe the issue
of quenching of the weak nucleon current operators in Section \ref{sec:quench}.

\begin{figure}[htb]
\centerline{\psfig{file=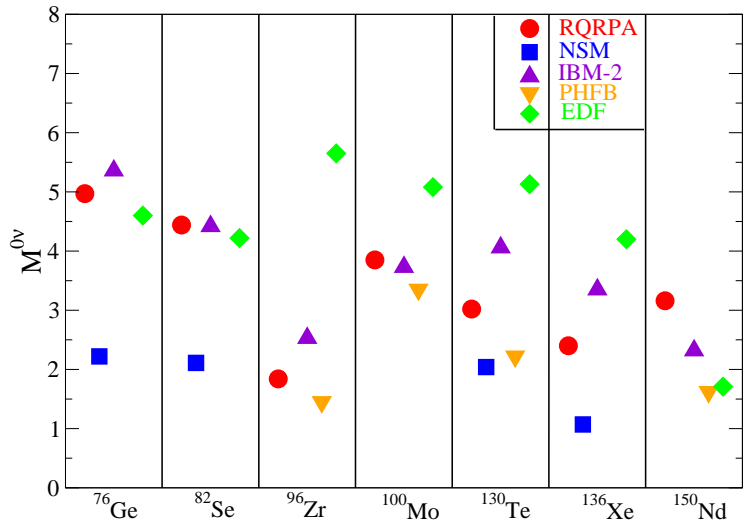 ,width=9.0cm}}
\caption{Dimensionless $0\nu\beta\beta$ nuclear matrix elements for selected nuclei
evaluated using a variety of indicated methods. For references see text.}
\label{fig:nme}
\end{figure}

The quasiparticle random phase approximation (QRPA) and its renormalized version
(RQRPA) is another method often used in the evaluation of $M^{0\nu}$. In it,
the valence space is not restricted and contains at least two full oscillator
shells, often more than that. On the other hand, only selected 
simple configurations of the valence nucleons are used. The basis states have
broken symmetries in which particle numbers, isospin, and possibly angular
momentum are not good quantum numbers but conserved only on average. After the equations of motion
are solved, some of the symmetries are partially restored. The RQRPA
partially restores the Pauli principle violation in the resulting
states.

The procedure consist of several steps. In the first one the like particle
pairing interaction is taken into account, using the BCS procedure. Then,
the neutron-proton interaction is used in the equations of motion, resulting
in states that contain two quasiparticle and two quasi hole configurations
and their iterations. Usually, the realistic G-matrix based interaction is used, 
but block renormalized with common renormalization factors. In particular,
all particle-particle channel interaction matrix elements are typically scaled by the
factor $g_{pp}$ with nominal, unrenormalized value $g_{pp}=1$.   In many applications,
including the one used in Figs.\ref{fig:nme}, \ref{fig:plotm} and \ref{fig:plott},
the parameter $g_{pp}$ is determined such that the experimental matrix element
$M^{2\nu}$ is correctly reproduced. This procedure has been first
suggested in Ref.\cite{Rod03}, resulting in $<$20\% change in $g_{pp}$.
 In this case, the method obviously is
not predicting the magnitude of $M^{2\nu}$, instead it uses its value
in the fit. However, as shown in \cite{Rod03}, the procedure removes
the dependence of $M^{0\nu}$ on the size of the valence space, making its
predicted value essentially constant. 
While most early QRPA and RQRPA calculations assume spherical
nuclear shape, recent Ref. \cite{Fang10} extension allows deformed 
nuclear shape. In Figs. 
\ref{fig:nme}, \ref{fig:plotm} and \ref{fig:plott}.
the RQRPA results are denoted by red circles.

The IBM-2 method uses the microscopic interacting boson model to
evaluate $M^{0\nu}$ \cite{Barea09}. In IBM-2 one begins with 
correlated $S$ and $D$ pairs of identical nucleons and includes
the effect of deformation through the bosonic neutron-proton
quadrupole interaction. The method describes well the low lying
states, the electromagnetic transitions between them and the two-nucleon
transition rates in spherical and strongly deformed nuclei. 
Even though the method was originally considered as an
approximation of the nuclear shell model, the resulting $M^{0\nu}$
(purple upward pointing triangles in 
Figs. \ref{fig:nme}, \ref{fig:plotm} and \ref{fig:plott}) are rather
surprisingly close to the RQRPA results and noticeably larger than the NSM ones.
At the present time it is not possible to evaluate the $M^{2\nu}$ matrix
elements within the IBM-2 method.

\begin{figure}
\centerline{\psfig{file=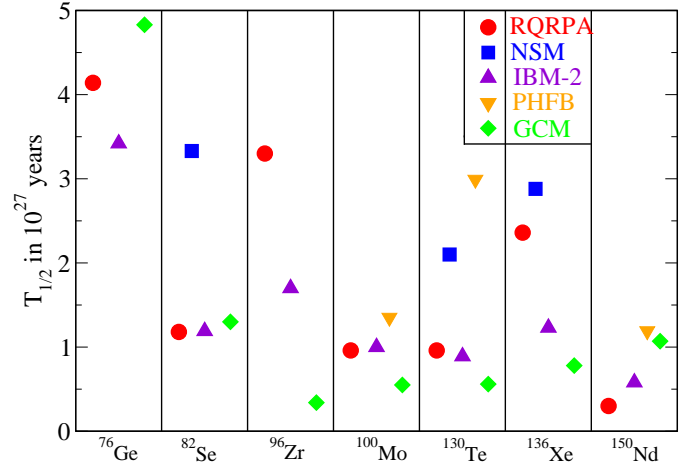 ,width=8.0cm}}
\caption{Halflives $T_{1/2}$ in units of $10^{27}$ years 
corresponding to the effective neutrino mass 
$< m_{\beta\beta} >$ = 20 meV and the matrix elements shown in Fig. \ref{fig:nme}. 
The lifetime $T_{1/2}$ scales as $< m_{\beta\beta} >^{-2}$.
Note that the NSM entry of 12.9$\times 10^{27}$ y for $^{76}$Ge is not shown.}
\label{fig:plotm}
\end{figure}
   
The projected Hartree-Fock-Bogoljubov framework \cite{Chat08} (PHFB) uses
angular momentum projected wave functions based on the hamiltonian with
pairing and quadrupole-quadrupole interaction. Again, the method is
not capable in describing the intermediate odd-odd nuclei and thus neither
the $M^{2\nu}$ matrix elements. The resulting $M^{0\nu}$ matrix elements for the
heavier nuclei (including the deformed $^{150}$Nd) are denoted with brown
downward pointing triangles in Figs. \ref{fig:nme}, \ref{fig:plotm} and \ref{fig:plott}.
This method, like IBM-2, does not explicitly contain the isoscalar neutron-proton
interaction, crucially important in QRPA. Yet its resulting $M^{0\nu}$ are in a
reasonable agreement with those from RQRPA.

\begin{figure}
\centerline{\psfig{file=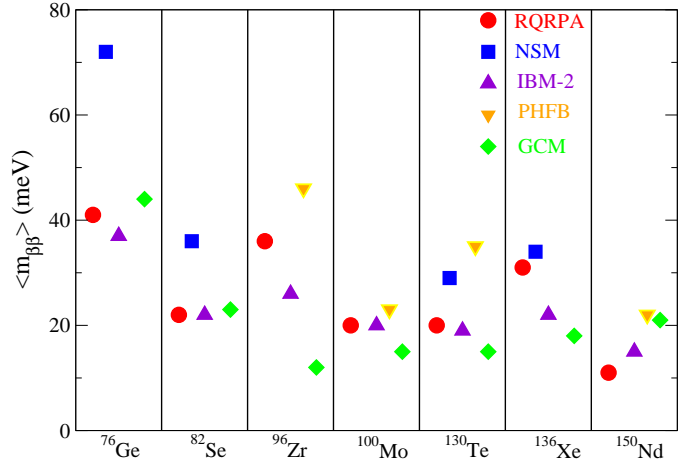,width=8.0cm}}
\caption{The effective neutrino mass $< m_{\beta\beta} >$ in units of meV evaluated assuming
that the halflife of all nuclei is $10^{27}$ years
and for the matrix elements shown in Fig. \ref{fig:nme}. Note that $< m_{\beta\beta} >$ scales
as $T_{1/2}^{-1/2}$.}
\label{fig:plott}
\end{figure}

In Ref. \cite{Rod10} the generating coordinate method (GCM) was used in conjunction with the
particle number and angular momentum projection for both the initial and final nuclei.
Large single particle space was used (11 shells) with the well established Gogny D1S
energy density functional. The initial and final many-body wave
functions are represented as combinations of the  particle number $N,Z$ projected
$I=0^+$ axially symmetric states with different intrinsic deformations. The lowest 
state is found by solving the Hill-Wheeler-Griffin eigenvalue equation. The resulting
$M^{0\nu}$ matrix elements are denoted with green diamonds in Figs.
 \ref{fig:nme},\ref{fig:plotm} and \ref{fig:plott}. Again, like in PHFB and IBM-2
methods, the treatment of the odd-odd nuclei, and thus also of the $M^{2\nu}$
matrix elements is impossible in GCM. 

Looking at the Fig.  \ref{fig:nme} one can see that, common to all five displayed
methods the predicted $M^{0\nu}$ nuclear matrix elements vary relatively smoothly,
with the mass number $A$, unlike the experimentally determined $M^{2\nu}$ matrix elements
displayed in Fig. \ref{fig_2nu} that vary strongly with $A$. 
The RQRPA, IBM-2, PHFB and GCM method are in a crude 
agreement with each other, and predict slow decrease of $M^{0\nu}$ with increasing $A$.
The $M^{0\nu}$ evaluated in NSM are essentially constant with $A$ and noticeably
smaller than those from the other methods, particularly in the lighter nuclei.
In Figs. \ref{fig:plotm} and \ref{fig:plott} I show the halflives $T_{1/2}$
for the fixed effective neutrino Majorana mass $\langle m_{\beta\beta} \rangle$
and   $\langle m_{\beta\beta} \rangle$ for fixed  $T_{1/2}$. These figures should
be helpful for comparing the experimental results 
and/or sensitivities for different nuclei.

The question that is, for obvious reasons, often asked is what is the error
or uncertainty of individual nuclear matrix elements? It is notoriosly
difficult to estimate the error of a theoretical result. One possibility
is just to use, without selection, all calculated values and take as the error
their spread. That was used in the provocative Ref. \cite{Bah04}. Clearly that is
not the correct approach, even though the quoted paper served as appropriate 
warning to the nuclear structure community.

Another possibility is to use as a measure of uncertainty the spread of selected
careful calculations, like those in Fig. \ref{fig:nme}. That also has problems.
Different methods use different approximations, and it is unlikely that all of
them are equally important. And it is difficult to decide which of them is
more realistic and which is less realistic.

Within a given nuclear structure method it is possible to assign a measure of error
by considering the uncertainties in the input parameters. That was done for QRPA
e.g. in Refs. \cite{Rod03,anatomy} where the size of the single particle basis,
whether QRPA or RQRPA was used, and whether $g_A$ was quenched or not is used for 
such estimate. Clearly that error does not include the possible systematic 
uncetainty of the basic method itself, but serves as a useful estimate of the irreducible
uncertainty. When considering the ratios of the matrix elements for different nuclei,
the question of correlation of the corresponding errors arises. As shown in Ref. \cite{Fogli} 
at least in the case of QRPA, the errors are highly correlated.

\section{$0\nu\beta\beta$ nuclear matrix elements: Physics considerations}

Double beta decay in both $2\nu$ and $0\nu$ modes can exist because even-even nuclei
are more bound than the neighboring odd-odd nuclei. This extra binding
is a consequence of {\it pairing} between like nucleons. In nonmagic systems
neutrons and/or protons form $0^+$ pairs and the corresponding Fermi level becomes
diffuse over the region  with the characteristic size $\sim$ pairing gap $\Delta$.
This opens more possibilities for $nn \rightarrow pp$ transitions. The calculated
matrix elements $M^{0\nu}$ increase when the gap $\Delta$ increases. The 
(unrealistic in real nuclei) situation of pure paired nuclear system (only seniority 0
states) would have very large $M^{0\nu}$.

\begin{figure}
\centerline{\psfig{file=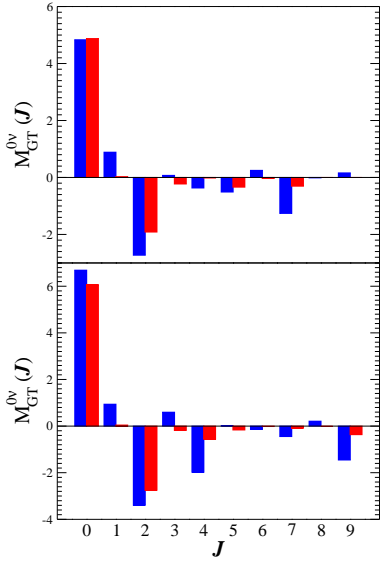,width=5.0cm}}
\caption{Contributions of different angular moments $\cal{J}$ of the two decaying neutrons
 to $M^{0\nu}_{GT}$ in $^{82}$Se (upper panel) and $^{130}$Te (lower panel).
 The results of NSM (red) and QRPA (blue) are compared. 
Both calculations use identical limited single particle spaces. 
The values of  $M^{0\nu}_{GT}$ are 1.32(QRPA) and 2.06(NSM) for $^{82}$Se
and 1.05(QRPA) and 1.98(NSM) for $^{130}$Te.} 
\label{fig:pair1}
\end{figure}

However, in real nuclei opposite tendencies exist. Real nuclei have admixtures of 
the ``broken pair'' states, or in the shell model language, states with higher
seniority. These states are present because other parts of the nucleon interaction
exist, in particular the neutron-proton force. It is illustrative to characterize 
such states by the angular momentum $\cal{J}$ of the neutron pair that is in the
$\beta\beta$ decay transformed into the proton pair with the same $\cal{J}$.
In Fig. \ref{fig:pair1} the corresponding competition is illustrated, 
in both NSM and QRPA. While the positive pairing parts are very similar (since
the same single particle spaces are used), the negative $\cal{J} \ne$ 0 parts
are only qualitatively similar. Nevertheless the severe cancellation between these
two tendencies, and between the corresponding components of the residual interaction,
is present in both methods.

\begin{figure}
\centerline{\psfig{file=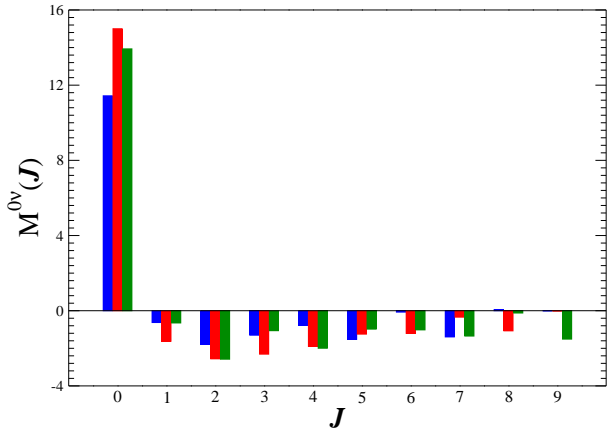,width=8.0cm}}
\caption{Contributions of different angular moments $\cal{J}$ of the two decaying neutrons
to $M^{0\nu}_{GT}$ in $^{76}$Ge (blue), $^{100}$Mo (red) and $^{130}$Te (green) 
within QRPA with proper single particle space
(two full oscillator shells). 
The $M^{0\nu}$ values are 3.98 ($^{76}$Ge), 2.74 ($^{100}$Mo and 2.67 ($^{130}$Te. }
\label{fig:pair2}
\end{figure}

The situation shown in Fig. \ref{fig:pair1} is somewhat unrealistic as far as the QRPA
is concerned; the corresponding single particle space is just too small. When it is enlarged
to contain two full oscillator shells, the competition becomes even more severe. Both
the $\cal{J}$ = 0 and  $\cal{J} \ne$ 0 grow, while their difference remains more or
less unchanged. This situation, illustrated in Fig.\ref{fig:pair2}, appears to be is universal.
One can see that, due to this cancellation, accurate evaluation of $M^{0\nu}$ is difficult;
small uncertainty in either of the two (positive and negative) components is strongly
enhanced in the difference.

The $0\nu\beta\beta$ decay operators depend on the internucleon distance $r_{12}$ 
due to the neutrino potential $H(r,\bar{E})$. Obviously, the range of $r_{12}$ is restricted
from above by $r_{12} \le 2R$. From the  form of $H(r) \sim R/r$ one could, naively,
expect that the characteristic value of $r_{12}$ is the typical distance between nucleons
in the nucleus, namely $\bar{r}_{12} \sim R$. However, that is not true, in reality only
much smaller values of $r_{12} \le$ 2-3 fm or equivalently larger values of the
momentum transfer $q$ are relevant. That was first demonstrated within QRPA in Ref. \cite{anatomy},
very similar result was obtained also in NSM \cite{Men09}.

To see how that conclusion is obtained, define a function $C(r)$
\begin{equation}
C^{0\nu}_{GT}(r) =  \langle f | \Sigma_{lk} \vec{\sigma}_l \cdot  \vec{\sigma}_k \tau_l^+ \tau_k^+
\delta(r - r_{lk}) H(r_{lk},\bar{E}) | i \rangle ~,
\label{eq:C(r)}
\end{equation}
Obviously, this function is normalized by
\begin{equation}
 M^{0\nu}_{GT} = \int_0^{\infty} C^{0\nu}_{GT} (r) dr ~,
 \label{eq:C(r)int}
\end{equation}
and is related to the function $C_{cl}^{2\nu}(r)$ (see Eq. (\ref{eq:2nuclp})) by
\begin{equation}
C^{0\nu}_{GT}(r) =  H(r,\bar{E}) \times C^{2\nu}_{cl}(r) ~,
\label{eq:basic}
\end{equation}
which is valid for any shape of the neutrino potential.

\begin{figure}
\centerline{\psfig{file=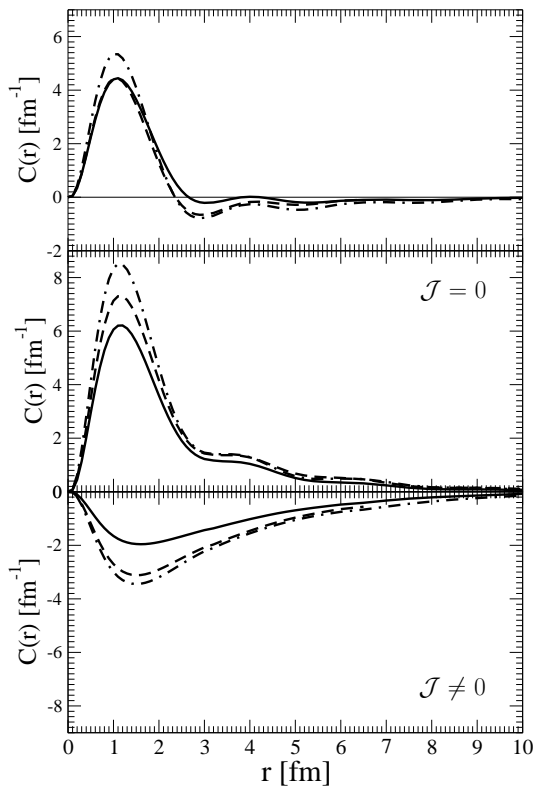,width=7.0cm}}
\caption{Dependence on $r_{12}$ for $^{76}$Ge (full line), 
$^{100}$Mo (dot-dashed line) and $^{130}$Te (dashed line)
evaluated in QRPA. The upper panel shows the full matrix element, and the lower
panel shows separately the ``pairing'' ($\cal{J}$ = 0) and ``broken pair''
 ($\cal{J} \ne$ 0) contributions. The integrated $M^{0\nu}$ are 5.4, 4.5, and 4.1
for these  three nuclei.}
\label{fig:radial}
\end{figure}

Examples of the functions $C^{0\nu}(r)$ are shown in Fig. \ref{fig:radial} for
three representative nuclei. As the lower panel demonstrates, the cancellation
between the ``pairing'' ($\cal{J}$ = 0) and ``broken pairs'' ($\cal{J} \ne$ 0) is 
essentially complete for $r_{12} \ge$ 2-3 fm (similar figure appears in \cite{anatomy}). 
That cancellation is so complete
when the particle-particle renormalization constant $g_{pp}$ is chosen in the
usual way, i.e. so that the $2\nu\beta\beta$ lifetime is correctly reproduced.
For other values of $g_{pp}$ the cancellation between $\cal{J}$ = 0 and $\cal{J} \ne$ 0
is less perfect. Analogous conclusions can be obtained in an exactly solvable
model \cite{EV04} based on the algebra $SO(5)\times SO(5)$.

The relation between $C^{0\nu}(r)$ and  $C^{2\nu}_{cl}(r)$ in Eq. (\ref{eq:basic})
is intriguing. Can it be used for the evaluation of $M^{0\nu}$? There are two problems
with that possibility. While $M^{2\nu}$ can be deduced from the measured lifetime of
the $2\nu\beta\beta$ decay, the closure matrix element cannot be deduced that way.
Moreover, even if it would be possible to somehow determine  $M^{2\nu}_{cl}$
accurately, its magnitude is insufficient to determine the whole function
 $C^{2\nu}_{cl}(r)$.

\section{Quenching of the weak axial current}
\label{sec:quench}

It is well known that experimental  Gamow-Teller $\beta$-decay transitions to individual final states
are noticeably weaker than the theory predicts. That phenomenon is known
as the axial current matrix elements {\it quenching}. The $\beta$-strength functions can
be studied also with the charge exchange nuclear reactions and a similar effect is
observed as well. Thus, in order to describe the matrix elements of the operator $\sigma \tau$,
the empirical rule $(\sigma \tau)^2_{eff} \simeq 0.6 (\sigma \tau)^2_{model}$ is usually
used (see \cite{Ost92,Bro85,Cau94}). Since these operators accompany weak axial current,
it is convenient to account for such quenching by using an effective coupling constant
$g^{eff}_A \sim 1.0$ instead of the true value $g_A = 1.269$.

The evidence for quenching is restricted so far to the Gamow-Teller operator $\sigma \tau$
and relatively low-lying final states.
It is not known whether the other multipole operators associated with the weak axial current
should be quenched as well. In fact, the analysis of the muon capture rates in
\cite{Kol00} and of the unique second forbidden $\beta$ decays in \cite{MV98}
suggests that quenching is not needed in those cases.

The quenching of the  axial current matrix elements, or simply of $g_A$, is believed to
be caused by nucleon correlations \cite{BH82}. The alternative explanation, screening 
of the GT operator by the $\Delta$-hole pairs \cite{Er73} is, these days, not considered
as the likely explanation. Very recently, in Ref. \cite{Men11}, the quenching has been
associated with the two-body  currents.

Since the $2\nu\beta\beta$ decay involves only the GT operators and relatively low-lying
intermediate states, one could expect that the quenching is involved in that case.
Indeed, as already mentioned, in the nuclear shell model the agreement with experimental
decay rate is achieved only with quenching  
$(\sigma \tau)^2_{eff} \simeq (0.2-0.36) (\sigma \tau)^2_{model}$,
with similar quenching required to reproduce the measured $\beta$ decay rates and the $\beta$
strength functions. 

In QRPA and RQRPA the usual way of the $g_{pp}$ renormalization using the experimental
$M^{2\nu}$ means that the effect of quenching cannot be deduced from the comparison
of the calculated and empirical $M^{2\nu}$. In the other discussed methods (IBM-2, PHFB, GCM)
it is not possible at the present time to calculate $M^{2\nu}$, so the question of quenching
for the $2\nu\beta\beta$ is meaningless in that case.

However, whether quenching should be applied to the $0\nu\beta\beta$ matrix elements $M^{0\nu}$
is an important issue and a source of noticeable uncertainty. Within QRPA and RQRPA the effect
of quenching on the  $0\nu\beta\beta$ decay mode has been considered in Refs. \cite{5th,Sim09,Rod06}.
The calculations there uses a modified definition 
\begin{equation}
M'^{0\nu} = \left(\frac{g^{eff}_A}{1.269}\right)^2 M^{0\nu}(g^{eff}_A),
\end{equation}
and thus the standard phase space factor can be used independently of whether quenching is
included or not. It is concluded there that the matrix elements $M'^{0\nu}$ are reduced by
$\sim$20-30\% when $g^{eff}_A = 1.0$ is assumed compared with $g_A = 1.269$. Thus, in that case
the  predicted $0\nu$ decay rates are affected by the possible quenching
less than naive expectation based om the ratio $[g^{eff}_A/g_A]^4$ might suggest.

Somewhat similar, but from the point of view of physics involved quite different 
reasons was reached in Ref. \cite{Men11} when the effect of two-body current is 
included in the effective field theory.

The problem of quenching of the $M^{0\nu}$ matrix elements remains one of
 the main sources of uncertainty
in their true value. The relative change with $A$ and $Z$, as depicted in 
Fig. \ref{fig:nme} will be,
however, affected only little by the inclusion of the quenching phenomenon.

\section{Conclusions}

In the last few years the problem of evaluation of the nuclear matrix elements $M^{0\nu}$ of 
$0\nu\beta\beta$ decay has entered a new era. As described here, there are now 
five or more seemingly quite different methods of solving it; it became one
of the very active subfields of nuclear structure theory. The results are
not yet quite convergent, but at least they suggest that the calculated values
are fairly insensitive (within a factor of about two) to the broad range
of approximations made. Moreover, all calculations agree that the values of $M^{0\nu}$
do not change abruptly from one candidate nucleus to another one. Thus, if the
$0\nu\beta\beta$ decay were observed in one nucleus, one can with some confidence
predict its lifetime in the other candidate nuclei, increasing the chances that
a reliable and confirmed result is obtained. Of course, the fact that the nuclear shell
model consistently predicts $M^{0\nu}$ values that are noticeably smaller
than in the other methods remains. Is it the inclusion of complicated configurations
(states with high seniority) that causes the reduction of $M^{0\nu}$ in NSM, or
is it the inclusion of larger single-particle spaces in most other methods?
The answer to this question is obviously needed.

Moreover, there are several more general issues that deserve 
closer attention of nuclear theorists.
One of them is the problem of quenching of the axial weak current matrix elements.
That effect is well established in ordinary allowed beta decays, and in the evaluation
of the beta strength functions involving low-lying nuclear states. Should an analogous
effect be applied to the theoretical evaluation of $M^{0\nu}$? This is, as yet,
an open question. Its solution could affect the true values of $M^{0\nu}$ appreciably,
by about 30\% according to our estimates.

Problem of quenching is just a part of the determination of effective operators.
How does one consistently include the renormalization caused by using only a subset
of the general Hilbert space in the calculational framework? Renormalization of the $g_A$,
i.e. inclusion of quenching, is an example of such effect. The problem of taking
into account the high momentum part of the nucleon interaction, causing the short
range repulsion, belongs to that category as well. There is a consensus now, that
the recent developments (see Section \ref{sect:short}) point us in the right direction.
More work in that direction would be clearly beneficial.

Finally, most attention up till now was concentrated on the so-called standard scenario,
according to which the $0\nu\beta\beta$ decay would be caused by the exchange of the light
Majorana neutrinos that interact through the canonical left-handed weak currents. Much less
attention has been given to the possible mechanism involving heavy, $\sim$ TeV, particle
exchange and thus extremely short range effects. Is it really generally true that in
these cases the pionic effects, described in Section \ref{sect:heavy}, dominate? If that
is the case much more detailed evaluations of the $M^{0\nu}$ in such cases is clearly needed.
In particular, in the next few years much progress, one hopes, in the exploration of the TeV
mass range particles will be achieved at LHC. If some of the suggested particle physics
models (Left-Right symmetry, R-parity violating supersymmetry) might find support
at LHC, more work on the corresponding $M^{0\nu}$ matrix elements will be clearly
needed.

\section{Acknowledgements}
The original results reported here were obtained in collaboration with
 Jonathan Engel, Amand Faessler, 
 Gary Prezeau,
 Michael Ramsey-Musolf, Vadim Rodin,  and Fedor \v{S}imkovic.
 The fruitful collaboration with them
 is gratefully acknowledged. The work was supported in part by the
US NSF Grant 0855538.

\bigskip

{\bf Appendix: Possible resonances in $0\nu ECEC$ decays}

The two electron capture decay without neutrino emission requires a special
comment. Clearly, when the initial and final states have different energies, the process
cannot proceed since energy is not conserved. The radiative process,
with bremsstrahlung photon emission, however,
can proceed and its rate, unlike all the other neutrinoless processes, increases
with decreasing $Q$ value \cite{SW04}. (However, the estimated decay rates are
quite small and lifetimes long).
In the extreme case of essentially perfect degeneracy the photon emission
is not needed, and a resonance
enhancement can occur \cite{Ber83}.

In the case of resonance 
the initial state is the atom $(Z,A)$, stable against ordinary $\beta$ decay. 
The final state
is the ion $(Z-2,A)$ with electron vacancies $H,H'$ and, in general, with the nucleus
in some excited state of energy $E^*$. The resonance occurs if the final energy
$E = E^* + E_H + E_{H'}$
is close to the decay $Q$ value,
i.e. the difference of the initial and final atomic masses,
and a perfect resonance occurs when $Q-E$ is less than
the width of the final state which is dominated by the electron hole widths 
$\Gamma_H, \Gamma_{H'}$.
The decay rate near resonance is given by the Breit-Wigner type formula
\begin{equation}
\frac{1}{\tau} = \frac{(\Delta M)^2}{(Q-E)^2 + \Gamma^2/4}\Gamma ~,
\end{equation}
where $\Delta M$ is the matrix element of weak interaction between the two degenerate
atomic states.

The states of definite energy, the eigenstates of the total hamiltonian, are superpositions
of the initial and final states, mixed by  $\Delta M$. But in reality, the initial state 
is pure, and not a state of definite energy, since the final state decays essentially  
immediately.

The mixing matrix element is \cite{Ber83}
\begin{equation}
\Delta M \sim  \frac{G_F^2 \cos^2 \theta_C}{4 \pi}  \langle m_{\beta \beta} \rangle
|\psi(0)|^2 g_A^2 M^{0\nu} ~,
\end{equation}
where $\psi(0)$ is the amplitude at the origin of the wave function of the
captured electrons and $M^{0\nu}$ is the nuclear matrix element, same one as before. 
Clearly, if the resonance can be approached, the decay rate would be
enhanced by the factor $4/\Gamma$
compared to $\Gamma/(E-Q)^2$, where
the width $\Gamma$ is typically tens of eV.
Estimates suggest that in such a case the decay lifetime for
$\langle m_{\beta \beta} \rangle \sim$ 1 eV could be of the order of
$10^{24-25}$ years. However, chances of finding
a case of a perfect (eV size) resonance when $E$ is of order of MeV are
not large. Indeed, in the best case found so far, in $^{152}$Gd \cite{Elis11},
the quantity $Q-E$ = 0.91(18) keV, still with the predicted halflife 
of only $\sim 10^{26}$ y
for $\langle m_{\beta\beta} \rangle$ = 1 eV.

\end{document}